%% file: 19SSCL_LAMA.tex
\documentclass[conference,10pt,twoside,twocolumn]{IEEEtran}

\usepackage{mleftright}
\usepackage{cite}
\usepackage[T1]{fontenc}
\usepackage{graphicx}
\usepackage{mathtools}
\usepackage{amssymb}
\usepackage{amsmath}
\usepackage{amsthm}

\usepackage{paralist}
\usepackage{subfigure}
\usepackage{booktabs} %\toprule \midrule \bottomrule
\usepackage{algorithm}
\usepackage{multirow}
\usepackage{microtype}
\usepackage{tablefootnote}

\input{macros/vmr-symbols-vecbold}
\input{macros/standard-macros}

\input{macros/defs}

\usepackage{framed}

\usepackage{color}

\IEEEoverridecommandlockouts
\allowdisplaybreaks % THIS ALLOWS EQUATIONS TO BREAK THE PAGE

\newcommand{\revision}[1]{\textcolor{black}{#1}}

\safemath{\MRT}{\textrm{MRT}}
\safemath{\betamax}{\beta^\textnormal{max}}
\safemath{\tmax}{t_\textnormal{max}}
\safemath{\betamaxno}{\beta^\textnormal{max}}
\safemath{\betamin}{\beta^\textnormal{min}}
\safemath{\betaminno}{\beta^\textnormal{min}}

\safemath{\Nomin}{\No^\textnormal{min}(\beta)}
\safemath{\Nominnobeta}{\No^\textnormal{min}}
\safemath{\Nomax}{\No^\textnormal{max}(\beta)}
\safemath{\Nomaxnobeta}{\No^\textnormal{max}}

\newcommand{\ProbMacro}[1]{\textnormal{P}\!\left[#1\right]}

\safemath{\shat}{\hat\bms}
\safemath{\shatelem}{\hat{s}}
\safemath{\sigmasqhat}{\hat\sigma^2}
\safemath{\sigmasq}{\sigma^2}
\safemath{\gammasqhat}{\hat\gamma^2}
\safemath{\setOreal}{\setO^\textnormal{real}}
\safemath{\setOimag}{\setO^\textnormal{imag}}
\safemath{\rhotilde}{\tilde{\rho}}
\safemath{\dd}{\textnormal{d}}

\safemath{\MAP}{\textrm{MAP}}
\safemath{\IO}{\textrm{IO}}
\safemath{\JO}{\textrm{JO}}
\safemath{\Nopost}{N_{0}^\textnormal{post}}

\safemath{\MT}{U}
\safemath{\mt}{u}
\safemath{\MR}{B}
\safemath{\mr}{b}
\safemath{\TsNumber}{36}
\safemath{\Ts}{T_\textnormal{s}}
\safemath{\TLLR}{T_\textnormal{LLR}}
\safemath{\areal}{a^\textnormal{R}}
\safemath{\sreal}{s^\textnormal{R}}
\safemath{\xreal}{x^\textnormal{R}}
\safemath{\zreal}{z^\textnormal{R}}

\safemath{\aimag}{a^\textnormal{I}}
\safemath{\simag}{s^\textnormal{I}}
\safemath{\ximag}{x^\textnormal{I}}
\safemath{\zimag}{z^\textnormal{I}}

\safemath{\bmalpha}{\boldsymbol{\alpha}}
\safemath{\bmrho}{\boldsymbol{\rho}}
\safemath{\bmtau}{\boldsymbol{\tau}}

\safemath{\LA}{\Lprior}
\safemath{\Lprior}{\Lambda^\textnormal{prior}}
\safemath{\bmLprior}{\bm{\Lambda}^\textnormal{prior}}
\safemath{\LABold}{\bm{\Lambda}^\textnormal{a}}
\safemath{\LE}{\Lambda^\textnormal{e}}
\safemath{\LD}{\Lambda^\textnormal{d}}
\safemath{\LDbold}{\bm{\Lambda}^\textnormal{d}}

\newcommand*{\fancyrefremarklabelprefix}{remark}
\frefformat{vario}{\fancyrefremarklabelprefix}{Remark~#1}

\safemath{\tildebG}{\widetilde{\bG}}
\safemath{\bmymf}{\bmy^\textnormal{MF}}
\safemath{\tildebmymf}{\widetilde{\bmy}^\textnormal{MF}}
\safemath{\diag}{\textnormal{diag}}

\safemath{\Tran}{\textnormal{T}}
\safemath{\Herm}{\textnormal{H}}
\safemath{\row}{\textnormal{r}}
\safemath{\col}{\textnormal{c}}

\safemath{\imagText}{{\textnormal{Im}}}
\safemath{\realText}{{\textnormal{Re}}}

\safemath{\EX}{E_\textnormal{x}}
\safemath{\EXP}{\EX^\textnormal{p}}

\usepackage{algcompatible}
\algrenewcommand\algorithmicindent{1em}

\usepackage{array}
\newcolumntype{P}[1]{>{\centering\arraybackslash}p{#1}}

\begin{document}

\title{
A 354\,Mb/s 0.37\,mm$^{\bf 2}$ 151\,mW 32-User 256-QAM Near-MAP Soft-Input Soft-Output Massive MU-MIMO Data Detector in 28nm CMOS}\author{\IEEEauthorblockN{Charles Jeon, Oscar Casta\~neda, and Christoph Studer}\\[-0.6cm]
% \\[-0.2cm]
\thanks{CJ, OC, and CS are with the School of ECE, Cornell University, Ithaca, NY; e-mail: studer@cornell.edu; web: http://vip.ece.cornell.edu.
The work was supported by Xilinx Inc.~and by the US NSF under grants ECCS-1408006, CCF-1535897,  CCF-1652065, CNS-1717559, and ECCS-1824379.}\thanks{The authors thank A.~Maleki for discussions on approximate message passing and F.~K.~G\"urkaynak for his assistance during ASIC testing.}}
% \vspace{-5cm}
\maketitle

% TRICK TO MAKE STUFF COMPACT
%\linespread{0.99}
\fussy

\begin{abstract}
This paper presents a novel data detector ASIC for massive multiuser multiple-input multiple-output (MU-MIMO) wireless systems.
The ASIC implements a modified version of the large-MIMO approximate message passing algorithm (LAMA), which achieves near-optimal error-rate performance (i) under realistic channel conditions and (ii) for systems with as many users as base-station (BS) antennas.
The hardware architecture supports 32 users transmitting up to 256-QAM simultaneously and in the same frequency band, and provides soft-input soft-output capabilities for iterative detection and decoding.
The fabricated 28nm CMOS ASIC occupies 0.37\,mm$^\text{2}$, achieves a throughput of 354\,Mb/s, consumes 151\,mW, and improves the SNR by more than 11\,dB compared to existing data detectors in systems with 32 BS antennas and 32 users for realistic wireless channels. 
In addition, the ASIC achieves 4$\boldsymbol\times$ higher throughput per area than a recently proposed message-passing detector.
\end{abstract}

\input{sec1-intro}

\input{sec2-system}
\input{sec4-vlsi.tex}

\input{sec5-implementation.tex}

\appendices
\input{secc-appendix.tex}

\bibliographystyle{IEEEtran}
\bibliography{bib/VIPabbrv,bib/confs-jrnls,bib/publishers,bib/cjBibTeX_190331}

\end{document}

%% file: macros/vmr-symbols-vecbold.tex
\usepackage{amssymb}
\usepackage{amsfonts}
\usepackage{mathrsfs}
\usepackage{xspace}
\usepackage{bm}
\usepackage{upgreek}

\newcommand{\safemath}[2]{\newcommand{#1}{\ensuremath{#2}\xspace}}

\safemath{\bma}{\mathbf{a}}
\safemath{\bmb}{\mathbf{b}}
\safemath{\bmc}{\mathbf{c}}
\safemath{\bmd}{\mathbf{d}}
\safemath{\bme}{\mathbf{e}}
\safemath{\bmf}{\mathbf{f}}
\safemath{\bmg}{\mathbf{g}}
\safemath{\bmh}{\mathbf{h}}
\safemath{\bmi}{\mathbf{i}}
\safemath{\bmj}{\mathbf{j}}
\safemath{\bmk}{\mathbf{k}}
\safemath{\bml}{\mathbf{l}}
\safemath{\bmm}{\mathbf{m}}
\safemath{\bmn}{\mathbf{n}}
\safemath{\bmo}{\mathbf{o}}
\safemath{\bmp}{\mathbf{p}}
\safemath{\bmq}{\mathbf{q}}
\safemath{\bmr}{\mathbf{r}}
\safemath{\bms}{\mathbf{s}}
\safemath{\bmt}{\mathbf{t}}
\safemath{\bmu}{\mathbf{u}}
\safemath{\bmv}{\mathbf{v}}
\safemath{\bmw}{\mathbf{w}}
\safemath{\bmx}{\mathbf{x}}
\safemath{\bmy}{\mathbf{y}}
\safemath{\bmz}{\mathbf{z}}
\safemath{\bmzero}{\mathbf{0}}
\safemath{\bmone}{\mathbf{1}}

\bmdefine{\biad}{a}
\bmdefine{\bibd}{b}
\bmdefine{\bicd}{c}
\bmdefine{\bidd}{d}
\bmdefine{\bied}{e}
\bmdefine{\bifd}{f}
\bmdefine{\bigd}{g}
\bmdefine{\bihd}{h}
\bmdefine{\biid}{i}
\bmdefine{\bijd}{j}
\bmdefine{\bikd}{k}
\bmdefine{\bild}{l}
\bmdefine{\bimd}{m}
\bmdefine{\bind}{n}
\bmdefine{\biod}{o}
\bmdefine{\bipd}{p}
\bmdefine{\biqd}{q}
\bmdefine{\bird}{r}
\bmdefine{\bisd}{s}
\bmdefine{\bitd}{t}
\bmdefine{\biud}{u}
\bmdefine{\bivd}{v}
\bmdefine{\biwd}{w}
\bmdefine{\bixd}{x}
\bmdefine{\biyd}{y}
\bmdefine{\bizd}{z}

\bmdefine{\bixid}{\xi}
\bmdefine{\bilambdad}{\lambda}
\bmdefine{\bimud}{\mu}
\bmdefine{\bithetad}{\theta}
\bmdefine{\biphid}{\phi}
\bmdefine{\bideltad}{\delta}

\safemath{\bmia}{\biad}
\safemath{\bmib}{\bibd}
\safemath{\bmic}{\bicd}
\safemath{\bmid}{\bidd}
\safemath{\bmie}{\bied}
\safemath{\bmif}{\bifd}
\safemath{\bmig}{\bigd}
\safemath{\bmih}{\bihd}
\safemath{\bmii}{\biid}
\safemath{\bmij}{\bijd}
\safemath{\bmik}{\bikd}
\safemath{\bmil}{\bild}
\safemath{\bmim}{\bimd}
\safemath{\bmin}{\bind}
\safemath{\bmio}{\biod}
\safemath{\bmip}{\bipd}
\safemath{\bmiq}{\biqd}
\safemath{\bmir}{\bird}
\safemath{\bmis}{\bisd}
\safemath{\bmit}{\bitd}
\safemath{\bmiu}{\biud}
\safemath{\bmiv}{\bivd}
\safemath{\bmiw}{\biwd}
\safemath{\bmix}{\bixd}
\safemath{\bmiy}{\biyd}
\safemath{\bmiz}{\bizd}

\safemath{\bmxi}{\bixid}
\safemath{\bmlambda}{\bilambdad}
\safemath{\bmmu}{\bimud}
\safemath{\bmtheta}{\bithetad}
\safemath{\bmphi}{\biphid}
\safemath{\bmdelta}{\bideltad}

\safemath{\bA}{\mathbf{A}}
\safemath{\bB}{\mathbf{B}}
\safemath{\bC}{\mathbf{C}}
\safemath{\bD}{\mathbf{D}}
\safemath{\bE}{\mathbf{E}}
\safemath{\bF}{\mathbf{F}}
\safemath{\bG}{\mathbf{G}}
\safemath{\bH}{\mathbf{H}}
\safemath{\bI}{\mathbf{I}}
\safemath{\bJ}{\mathbf{J}}
\safemath{\bK}{\mathbf{K}}
\safemath{\bL}{\mathbf{L}}
\safemath{\bM}{\mathbf{M}}
\safemath{\bN}{\mathbf{N}}
\safemath{\bO}{\mathbf{O}}
\safemath{\bP}{\mathbf{P}}
\safemath{\bQ}{\mathbf{Q}}
\safemath{\bR}{\mathbf{R}}
\safemath{\bS}{\mathbf{S}}
\safemath{\bT}{\mathbf{T}}
\safemath{\bU}{\mathbf{U}}
\safemath{\bV}{\mathbf{V}}
\safemath{\bW}{\mathbf{W}}
\safemath{\bX}{\mathbf{X}}
\safemath{\bY}{\mathbf{Y}}
\safemath{\bZ}{\mathbf{Z}}

\safemath{\bZero}{\mathbf{0}}
\safemath{\bOne}{\mathbf{1}}
\safemath{\bDelta}{\mathbf{\Delta}}
\safemath{\bLambda}{\mathbf{\UpLambda}}
\safemath{\bPhi}{\mathbf{\Upphi}}
\safemath{\bSigma}{\mathbf{\Upsigma}}
\safemath{\bOmega}{\mathbf{\Upomega}}
\safemath{\bTheta}{\mathbf{\Uptheta}}

\bmdefine{\biAd}{A}
\bmdefine{\biBd}{B}
\bmdefine{\biCd}{C}
\bmdefine{\biDd}{D}
\bmdefine{\biEd}{E}
\bmdefine{\biFd}{F}
\bmdefine{\biGd}{G}
\bmdefine{\biHd}{H}
\bmdefine{\biId}{I}
\bmdefine{\biJd}{J}
\bmdefine{\biKd}{K}
\bmdefine{\biLd}{L}
\bmdefine{\biMd}{M}
\bmdefine{\biOd}{N}
\bmdefine{\biPd}{O}
\bmdefine{\biQd}{P}
\bmdefine{\biRd}{R}
\bmdefine{\biSd}{S}
\bmdefine{\biTd}{T}
\bmdefine{\biUd}{U}
\bmdefine{\biVd}{V}
\bmdefine{\biWd}{W}
\bmdefine{\biXd}{X}
\bmdefine{\biYd}{Y}
\bmdefine{\biZd}{Z}

\bmdefine{\biDelta}{\Delta}
\bmdefine{\biLambda}{\Lambda}
\bmdefine{\biPhi}{\Phi}
\bmdefine{\biSigma}{\Sigma}
\bmdefine{\biOmega}{\Omega}
\bmdefine{\biTheta}{\Theta}

\safemath{\bimA}{\biAd}
\safemath{\bimB}{\biBd}
\safemath{\bimC}{\biCd}
\safemath{\bimD}{\biDd}
\safemath{\bimE}{\biEd}
\safemath{\bimF}{\biFd}
\safemath{\bimG}{\biGd}
\safemath{\bimH}{\biHd}
\safemath{\bimI}{\biId}
\safemath{\bimJ}{\biJd}
\safemath{\bimK}{\biKd}
\safemath{\bimL}{\biLd}
\safemath{\bimM}{\biMd}
\safemath{\bimN}{\biNd}
\safemath{\bimO}{\biOd}
\safemath{\bimP}{\biPd}
\safemath{\bimQ}{\biQd}
\safemath{\bimR}{\biRd}
\safemath{\bimS}{\biSd}
\safemath{\bimT}{\biTd}
\safemath{\bimU}{\biUd}
\safemath{\bimV}{\biVd}
\safemath{\bimW}{\biWd}
\safemath{\bimX}{\biXd}
\safemath{\bimY}{\biYd}
\safemath{\bimZ}{\biZd}

\safemath{\bimDelta}{\biDelta}
\safemath{\bimLambda}{\biLambda}
\safemath{\bimPhi}{\biPhi}
\safemath{\bimSigma}{\biSigma}
\safemath{\bimOmega}{\biOmega}
\safemath{\bimTheta}{\biTheta}

\safemath{\setA}{\mathcal{A}}
\safemath{\setB}{\mathcal{B}}
\safemath{\setC}{\mathcal{C}}
\safemath{\setD}{\mathcal{D}}
\safemath{\setE}{\mathcal{E}}
\safemath{\setF}{\mathcal{F}}
\safemath{\setG}{\mathcal{G}}
\safemath{\setH}{\mathcal{H}}
\safemath{\setI}{\mathcal{I}}
\safemath{\setJ}{\mathcal{J}}
\safemath{\setK}{\mathcal{K}}
\safemath{\setL}{\mathcal{L}}
\safemath{\setM}{\mathcal{M}}
\safemath{\setN}{\mathcal{N}}
\safemath{\setO}{\mathcal{O}}
\safemath{\setP}{\mathcal{P}}
\safemath{\setQ}{\mathcal{Q}}
\safemath{\setR}{\mathcal{R}}
\safemath{\setS}{\mathcal{S}}
\safemath{\setT}{\mathcal{T}}
\safemath{\setU}{\mathcal{U}}
\safemath{\setV}{\mathcal{V}}
\safemath{\setW}{\mathcal{W}}
\safemath{\setX}{\mathcal{X}}
\safemath{\setY}{\mathcal{Y}}
\safemath{\setZ}{\mathcal{Z}}
\safemath{\emptySet}{\varnothing}

\safemath{\colA}{\mathscr{A}}
\safemath{\colB}{\mathscr{B}}
\safemath{\colC}{\mathscr{C}}
\safemath{\colD}{\mathscr{D}}
\safemath{\colE}{\mathscr{E}}
\safemath{\colF}{\mathscr{F}}
\safemath{\colG}{\mathscr{G}}
\safemath{\colH}{\mathscr{H}}
\safemath{\colI}{\mathscr{I}}
\safemath{\colJ}{\mathscr{J}}
\safemath{\colK}{\mathscr{K}}
\safemath{\colL}{\mathscr{L}}
\safemath{\colM}{\mathscr{M}}
\safemath{\colN}{\mathscr{N}}
\safemath{\colO}{\mathscr{O}}
\safemath{\colP}{\mathscr{P}}
\safemath{\colQ}{\mathscr{Q}}
\safemath{\colR}{\mathscr{R}}
\safemath{\colS}{\mathscr{S}}
\safemath{\colT}{\mathscr{T}}
\safemath{\colU}{\mathscr{U}}
\safemath{\colV}{\mathscr{V}}
\safemath{\colW}{\mathscr{W}}
\safemath{\colX}{\mathscr{X}}
\safemath{\colY}{\mathscr{Y}}
\safemath{\colZ}{\mathscr{Z}}

\safemath{\opA}{\mathbb{A}}
\safemath{\opB}{\mathbb{B}}
\safemath{\opC}{\mathbb{C}}
\safemath{\opD}{\mathbb{D}}
\safemath{\opE}{\mathbb{E}}
\safemath{\opF}{\mathbb{F}}
\safemath{\opG}{\mathbb{G}}
\safemath{\opH}{\mathbb{H}}
\safemath{\opI}{\mathbb{I}}
\safemath{\opJ}{\mathbb{J}}
\safemath{\opK}{\mathbb{K}}
\safemath{\opL}{\mathbb{L}}
\safemath{\opM}{\mathbb{M}}
\safemath{\opN}{\mathbb{N}}
\safemath{\opO}{\mathbb{O}}
\safemath{\opP}{\mathbb{P}}
\safemath{\opQ}{\mathbb{Q}}
\safemath{\opR}{\mathbb{R}}
\safemath{\opS}{\mathbb{S}}
\safemath{\opT}{\mathbb{T}}
\safemath{\opU}{\mathbb{U}}
\safemath{\opV}{\mathbb{V}}
\safemath{\opW}{\mathbb{W}}
\safemath{\opX}{\mathbb{X}}
\safemath{\opY}{\mathbb{Y}}
\safemath{\opZ}{\mathbb{Z}}
\safemath{\opZero}{\mathbb{O}}
\safemath{\identityop}{\opI}

\safemath{\veca}{\bma}
\safemath{\vecb}{\bmb}
\safemath{\vecc}{\bmc}
\safemath{\vecd}{\bmd}
\safemath{\vece}{\bme}
\safemath{\vecf}{\bmf}
\safemath{\vecg}{\bmg}
\safemath{\vech}{\bmh}
\safemath{\veci}{\bmi}
\safemath{\vecj}{\bmj}
\safemath{\veck}{\bmk}
\safemath{\vecl}{\bml}
\safemath{\vecm}{\bmm}
\safemath{\vecn}{\bmn}
\safemath{\veco}{\bmo}
\safemath{\vecp}{\bmp}
\safemath{\vecq}{\bmq}
\safemath{\vecr}{\bmr}
\safemath{\vecs}{\bms}
\safemath{\vect}{\bmt}
\safemath{\vecu}{\bmu}
\safemath{\vecv}{\bmv}
\safemath{\vecw}{\bmw}
\safemath{\vecx}{\bmx}
\safemath{\vecy}{\bmy}
\safemath{\vecz}{\bmz}

\safemath{\veczero}{\bmzero}
\safemath{\vecone}{\bmone}
\safemath{\vecxi}{\bmxi}
\safemath{\veclambda}{\bmlambda}
\safemath{\vecmu}{\bmmu}
\safemath{\vectheta}{\bmtheta}
\safemath{\vecphi}{\bmphi}
\safemath{\vecdelta}{\bmdelta}

\safemath{\matA}{\bA}
\safemath{\matB}{\bB}
\safemath{\matC}{\bC}
\safemath{\matD}{\bD}
\safemath{\matE}{\bE}
\safemath{\matF}{\bF}
\safemath{\matG}{\bG}
\safemath{\matH}{\bH}
\safemath{\matI}{\bI}
\safemath{\matJ}{\bJ}
\safemath{\matK}{\bK}
\safemath{\matL}{\bL}
\safemath{\matM}{\bM}
\safemath{\matN}{\bN}
\safemath{\matO}{\bO}
\safemath{\matP}{\bP}
\safemath{\matQ}{\bQ}
\safemath{\matR}{\bR}
\safemath{\matS}{\bS}
\safemath{\matT}{\bT}
\safemath{\matU}{\bU}
\safemath{\matV}{\bV}
\safemath{\matW}{\bW}
\safemath{\matX}{\bX}
\safemath{\matY}{\bY}
\safemath{\matZ}{\bZ}
\safemath{\matzero}{\bmzero}

\safemath{\matDelta}{\bDelta}
\safemath{\matLambda}{\bLambda}
\safemath{\matPhi}{\bPhi}
\safemath{\matSigma}{\bSigma}
\safemath{\matOmega}{\bOmega}
\safemath{\matTheta}{\bTheta}

\safemath{\matidentity}{\matI}
\safemath{\matone}{\matO}

\safemath{\rnda}{A}
\safemath{\rndb}{B}
\safemath{\rndc}{C}
\safemath{\rndd}{D}
\safemath{\rnde}{E}
\safemath{\rndf}{F}
\safemath{\rndg}{G}
\safemath{\rndh}{H}
\safemath{\rndi}{I}
\safemath{\rndj}{J}
\safemath{\rndk}{K}
\safemath{\rndl}{L}
\safemath{\rndm}{M}
\safemath{\rndn}{N}
\safemath{\rndo}{O}
\safemath{\rndp}{P}
\safemath{\rndq}{Q}
\safemath{\rndr}{R}
\safemath{\rnds}{S}
\safemath{\rndt}{T}
\safemath{\rndu}{U}
\safemath{\rndv}{V}
\safemath{\rndw}{W}
\safemath{\rndx}{X}
\safemath{\rndy}{Y}
\safemath{\rndz}{Z}

\safemath{\rveca}{\bimA}
\safemath{\rvecb}{\bimB}
\safemath{\rvecc}{\bimC}
\safemath{\rvecd}{\bimD}
\safemath{\rvece}{\bimE}
\safemath{\rvecf}{\bimF}
\safemath{\rvecg}{\bimG}
\safemath{\rvech}{\bimH}
\safemath{\rveci}{\bimI}
\safemath{\rvecj}{\bimJ}
\safemath{\rveck}{\bimK}
\safemath{\rvecl}{\bimL}
\safemath{\rvecm}{\bimM}
\safemath{\rvecn}{\bimN}
\safemath{\rveco}{\bomO}
\safemath{\rvecp}{\bimP}
\safemath{\rvecq}{\bimQ}
\safemath{\rvecr}{\bimR}
\safemath{\rvecs}{\bimS}
\safemath{\rvect}{\bimT}
\safemath{\rvecu}{\bimU}
\safemath{\rvecv}{\bimV}
\safemath{\rvecw}{\bimW}
\safemath{\rvecx}{\bimX}
\safemath{\rvecy}{\bimY}
\safemath{\rvecz}{\bimZ}

\safemath{\rvecxi}{\bmxi}
\safemath{\rveclambda}{\bmlambda}
\safemath{\rvecmu}{\bmmu}
\safemath{\rvectheta}{\bmtheta}
\safemath{\rvecphi}{\bmphi}

\safemath{\rmatA}{\bimA}
\safemath{\rmatB}{\bimB}
\safemath{\rmatC}{\bimC}
\safemath{\rmatD}{\bimD}
\safemath{\rmatE}{\bimE}
\safemath{\rmatF}{\bimF}
\safemath{\rmatG}{\bimG}
\safemath{\rmatH}{\bimH}
\safemath{\rmatI}{\bimI}
\safemath{\rmatJ}{\bimJ}
\safemath{\rmatK}{\bimK}
\safemath{\rmatL}{\bimL}
\safemath{\rmatM}{\bimM}
\safemath{\rmatN}{\bimN}
\safemath{\rmatO}{\bimO}
\safemath{\rmatP}{\bimP}
\safemath{\rmatQ}{\bimQ}
\safemath{\rmatR}{\bimR}
\safemath{\rmatS}{\bimS}
\safemath{\rmatT}{\bimT}
\safemath{\rmatU}{\bimU}
\safemath{\rmatV}{\bimV}
\safemath{\rmatW}{\bimW}
\safemath{\rmatX}{\bimX}
\safemath{\rmatY}{\bimY}
\safemath{\rmatZ}{\bimZ}

\safemath{\rmatDelta}{\bimDelta}
\safemath{\rmatLambda}{\bimLambda}
\safemath{\rmatPhi}{\bimPhi}
\safemath{\rmatSigma}{\bimSigma}
\safemath{\rmatOmega}{\bimOmega}
\safemath{\rmatTheta}{\bimTheta}

%% file: macros/standard-macros.tex
\usepackage{amssymb}
\usepackage{amsfonts}
\usepackage{mathrsfs}
\usepackage{xspace}
\usepackage{bm}
\usepackage{fancyref}
\usepackage{textcomp}

\usepackage{multirow}
\usepackage{stmaryrd}

\newenvironment{textbmatrix}{	\setlength{\arraycolsep}{2.5pt}%
								\big[\begin{matrix}}{\end{matrix}\big]%
								\raisebox{0.08ex}{\vphantom{M}}}

\def\be{\begin{equation}}
\def\ee{\end{equation}}
\def\een{\nonumber \end{equation}}
\def\mat{\begin{bmatrix}}
\def\emat{\end{bmatrix}}
\def\btm{\begin{textbmatrix}}
\def\etm{\end{textbmatrix}}

\def\ba#1\ea{\begin{align}#1\end{align}}
\def\bas#1\eas{\begin{align*}#1\end{align*}}
\def\bs#1\es{\begin{split}#1\end{split}}
\def\bg#1\eg{\begin{gather}#1\end{gather}}
\def\bml#1\eml{\begin{multline}#1\end{multline}}
\def\bi#1\ei{\begin{itemize}#1\end{itemize}}

\DeclareMathOperator{\Exop}{\opE}			% expectation operator
\safemath{\dirac}{\delta}					% Dirac delta
\safemath{\krond}{\dirac}					% Kronecker delta
\safemath{\upto}{\uparrow}
\safemath{\downto}{\downarrow}
\safemath{\iu}{j}							% imaginary unit
\safemath{\ev}{\lambda}						% eigenvalue
\safemath{\hilseqspace}{l^{2}}				% Hilbert sequence space
\newcommand{\banachfunspace}[1]{\setL^{#1}}	% Banach function space
\safemath{\hilfunspace}{\banachfunspace{2}}	% Hilbert function space
\safemath{\SNR}{\textit{SNR}} 				% signal to noise ratio
\safemath{\PAR}{\textit{PAR}} 				% signal to noise ratio
\safemath{\No}{N_0}							% noise spectral density
\safemath{\Es}{E_s}							% energy per symbol
\safemath{\Eb}{E_b}							% energy per bit
\safemath{\EbNo}{\frac{\Eb}{\No}}
\safemath{\EsNo}{\frac{\Es}{\No}}

\DeclareMathOperator{\CHop}{\ensuremath{\opH}} % channel operator
\safemath{\tvir}{\rndh_{\CHop}}				% time-varying impulse response
\safemath{\tvtf}{\rndl_{\CHop}}				% 	-''- transfer function
\safemath{\spf}{\rnds_{\CHop}}				% spreading function
\safemath{\bff}{H_{\CHop}}					% bi-freuqency function

\safemath{\ircf}{r_{h}}						% impulse response correlation fn.
\safemath{\tftvcf}{r_{s}}					% scattering function
\safemath{\tfcf}{r_{l}}						% time-frequency correlation fn.
\safemath{\bfcf}{r_{H}}						% bi-frequency correlation fn.

\safemath{\tcorr}{c_h}						% time-correlation function
\safemath{\scf}{c_{s}}						% spreading function
\safemath{\tfcorr}{c_{l}}					% transfer-function correlation
\safemath{\fcorr}{c_{H}}						% frequency-correlation function

\safemath{\mi}{I}							% mutual information
\safemath{\capacity}{C}						% capacity

\safemath{\normal}{\mathcal{N}}			% normal distribution
\safemath{\jpg}{\mathcal{CN}}			% jointly proper Gaussian
\safemath{\mchain}{\leftrightarrow}		% Markov chain
\safemath{\dB}{\,\mathrm{dB}}
\safemath{\dBm}{\,\mathrm{dBm}}
\safemath{\Hz}{\,\mathrm{Hz}}
\safemath{\kHz}{\,\mathrm{kHz}}
\safemath{\MHz}{\,\mathrm{MHz}}
\safemath{\GHz}{\,\mathrm{GHz}}
\safemath{\s}{\,\mathrm{s}}
\safemath{\ms}{\,\mathrm{ms}}
\safemath{\mus}{\,\mathrm{\text{\textmu}s}}
\safemath{\ns}{\,\mathrm{ns}}
\safemath{\ps}{\,\mathrm{ps}}
\safemath{\meter}{\,\mathrm{m}}
\safemath{\mm}{\,\mathrm{mm}}
\safemath{\cm}{\,\mathrm{cm}}
\safemath{\m}{\,\mathrm{m}}
\safemath{\W}{\,\mathrm{W}}
\safemath{\mW}{\, \mathrm{mW}}
\safemath{\J}{\,\mathrm{J}}
\safemath{\K}{\,\mathrm{K}}
\safemath{\bit}{\,\mathrm{bit}}
\safemath{\nat}{\,\mathrm{nat}}

\safemath{\define}{\triangleq}			% definition

\safemath{\equivalent}{\sim}
\safemath{\distas}{\sim}					% distributed according to
\safemath{\sdiff}{\Delta}				% symmetric set difference

\safemath{\reals}{\mathbb{R}}
\safemath{\positivereals}{\reals_{+}}
\safemath{\integers}{\mathbb{Z}}
\safemath{\posint}{\integers_{+}}
\safemath{\naturals}{\mathbb{N}}
\safemath{\posnaturals}{\naturals_{+}}
\safemath{\complexset}{\mathbb{C}}
\safemath{\rationals}{\mathbb{Q}}

\newcommand*{\fancyrefapplabelprefix}{app}		% Appendix
\newcommand*{\fancyrefthmlabelprefix}{thm}		% Theorem
\newcommand*{\fancyreflemlabelprefix}{lem}		% Lemma
\newcommand*{\fancyrefcorlabelprefix}{cor}		% Corollary
\newcommand*{\fancyrefdeflabelprefix}{def}		% Definition
\newcommand*{\fancyrefproplabelprefix}{prop}		% Proposition
\newcommand*{\fancyrefexmpllabelprefix}{exmpl}
\newcommand*{\fancyrefalglabelprefix}{alg}		% Algorithm
\newcommand*{\fancyreftbllabelprefix}{tbl}		% Algorithm

\frefformat{vario}{\fancyrefseclabelprefix}{Section~#1}
\frefformat{vario}{\fancyrefthmlabelprefix}{Theorem~#1}
\frefformat{vario}{\fancyreftbllabelprefix}{Table~#1}
\frefformat{vario}{\fancyreflemlabelprefix}{Lemma~#1}
\frefformat{vario}{\fancyrefcorlabelprefix}{Corollary~#1}
\frefformat{vario}{\fancyrefdeflabelprefix}{Definition~#1}
\frefformat{vario}{\fancyreffiglabelprefix}{Fig.~#1}
\frefformat{vario}{\fancyrefapplabelprefix}{Appendix~#1}
\frefformat{vario}{\fancyrefeqlabelprefix}{(#1)}
\frefformat{vario}{\fancyrefproplabelprefix}{Proposition~#1}
\frefformat{vario}{\fancyrefexmpllabelprefix}{Example~#1}
\frefformat{vario}{\fancyrefalglabelprefix}{Algorithm~#1}

%% file: macros/defs.tex
\safemath{\dictab}{[\,\dicta\,\,\dictb\,]}

\safemath{\ysig}{\bmy}
\safemath{\ysighat}{\hat{\ysig}}
\safemath{\ysigdim}{M}
\safemath{\xsig}{\bmx}
\safemath{\xsigdim}{N}
\safemath{\nx}{n_x}
\safemath{\zsig}{\bmz}
\safemath{\zsigdim}{\ysigdim}
\safemath{\rsig}{\bmr}
\safemath{\Adict}{\bA}
\safemath{\Adicttilde}{\widetilde{\Adict}}
\safemath{\Adictdim}{\outputdim\times\xsigdim}
\safemath{\avec}{\bma}
\safemath{\avectilde}{\tilde{\avec}}
\safemath{\Bdict}{\bB}
\safemath{\Bdicttilde}{\widetilde{\Bdict}}
\safemath{\Cdict}{\bC}
\safemath{\cvec}{\bmc}
\safemath{\Ddict}{\bD}
\safemath{\Ddictdim}{\ysigdim\times\xsigdim}
\safemath{\dvec}{\bmd}
\safemath{\Ddicttilde}{\widetilde{\bD}}
\safemath{\Bonb}{\bB}
\safemath{\bvec}{\bmb}
\safemath{\Bonbdim}{\ysigdim\times\ysigdim}
\safemath{\noise}{\bmn}
\safemath{\noisedim}{\ysigim}
\safemath{\err}{\bme}
\safemath{\errdim}{\ysigdim}
\safemath{\errset}{\setE}
\safemath{\nerr}{n_e}
\safemath{\delop}{\bP_\errset}
\safemath{\delopc}{\bP_{{\errset}^c}}

\safemath{\cplxi}{\imath}
\safemath{\cplxj}{\jmath}
\safemath{\dict}{\matD}
\safemath{\inputdim}{N}		% number of columns of dictionary D
\safemath{\outputdim}{M}		%number of rows of dictionary D
\safemath{\sparsity}{S}	%sparsity
\safemath{\inputdimA}{{N_a}}	%total number of elements in dictionary A
\safemath{\inputdimB}{{N_b}}	%total number of elements in dictionary B
\safemath{\elemA}{{n_a}}	%number of elements chosen from dictionary A
\safemath{\elemB}{{n_b}}	%number of elements chosen from dictionary B
\safemath{\resA}{\matR_a}	%restriction map to elements of dictionary A
\safemath{\resB}{\matR_b}	%restriction map to elements of dictionary B
\safemath{\subD}{\matS} %subdictionary
\safemath{\subA}{\matS_a} %subdictionary part of A
\safemath{\subB}{\matS_b} %subdictionary part of B
\safemath{\dicta}{\matA} 	% first subdictionary
\safemath{\dictb}{\matB} 	% second subdictionary
\safemath{\hollowS}{H}
\safemath{\hollowA}{H_a}
\safemath{\hollowB}{H_b}
\safemath{\cross}{Z}
\safemath{\coh}{\mu_d}			% coherence dictionary
\safemath{\coha}{\mu_a}			% coherence first subdictionary
\safemath{\cohb}{\mu_b}			% coherence second subdictionary
\safemath{\mubs}{\nu}	%block sub-coherence
\safemath{\cohm}{\mu_m} %mutual coherence
\safemath{\dictset}{\setD}	% set of dictionaries
\safemath{\dictsetp}{\dictset(\coh,\coha,\cohb)}	% set of dictionaries parametrized
\safemath{\dictsetgen}{\dictset_\text{gen}}
\safemath{\dictsetgenp}{\dictsetgen(\coh)}
\safemath{\dictsetonb}{\dictset_\text{onb}}
\safemath{\dictsetonbp}{\dictsetonb(\coh)}

\safemath{\leftside}{U}
\safemath{\rightsideA}{R_a}
\safemath{\rightsideB}{R_b}

\safemath{\indexS}{\setI_S} %set of indices participating in sub-dictionary S

\safemath{\na}{n_a}			% cardinality of set of linearly independent columns of first dictionary
\safemath{\nb}{n_b}			% cardinality of set of linearly independent columns of second dictionary
\safemath{\coeffa}{p_i}	%coefficients for columns of A
\safemath{\coeffb}{q_j}	%coefficients for columns of B
\safemath{\seta}{\setP}		% set of linearly independent columns of A
\safemath{\setb}{\setQ}     % set of linearly independent columns of B
\safemath{\setw}{\setW}	%set of n largest elements of w
\safemath{\setz}{\setZ}	%set of L-n largest elements of z
\safemath{\cola}{\veca}		% generic element of the dictionary A
\safemath{\colb}{\vecb}		% generic element of the dictionary B
\safemath{\cold}{\vecd}		% generic element of the dictionary D
\safemath{\inputvec}{\vecx} 	%coefficient vector (input)
\safemath{\error}{\vece}	%error vector
\safemath{\noiseout}{\vecz} 	%noisy output vector
\safemath{\inputvecel}{x}
\safemath{\inputveca}{\vecx_a}
\safemath{\inputvecb}{\vecx_b}
\safemath{\outputvec}{\vecy}	%output of Dictionary
\safemath{\lambdamin}{\lambda_{\mathrm{min}}}
\safemath{\elltwo}{\ell_2}
\safemath{\ellone}{\ell_1}
\safemath{\ellzero}{\ell_0}
\safemath{\ellinf}{\ell_\infty}
\safemath{\ellinftilde}{\ell_{\widetilde\infty}}
\safemath{\licard}{Z(\coh,\coha,\cohb)}
\safemath{\xsol}{\hat{x}}
\safemath{\xbord}{x_b}		%Solution at the border
\safemath{\xstat}{x_s}		%Solution stationary in l0 prob
\safemath{\xstatLone}{\tilde{x}_s}
\safemath{\order}{\mathcal{O}} %order notation (big O)
\safemath{\scales}{\Theta} %scales as
\safemath{\ones}{\mathbf{1}} %all ones matrix
\safemath{\zeroes}{\mathbf{0}} %all zeroes matrix
\safemath{\thlone}{\kappa(\coh,\cohb)} %treshold l1 problem
\safemath{\constoneA}{\delta} %constant in l1 theorem to save space
\safemath{\constoneB}{\epsilon} %constant in l1 theorem to save space
\safemath{\nlarge}{L}				   %num large elements
\safemath{\sumlarge}{S_\nlarge}
\safemath{\maxlarger}{P_\nlarge}	   % maximum in Gribonval and Nielsen
\safemath{\Pzero}{\textrm{P0}}	
\safemath{\Pone}{\textrm{P1}}
\safemath{\vecfir}{\vecw}			 % \vecv element of the kernel of the dictionary, \vecv=[\vecfir \vecsec]
\safemath{\vecsec}{\vecz}
\safemath{\elvecfir}{w}              % element of vecfir
\safemath{\elvecsec}{z}				 % element of vecsec
\safemath{\nlargefir}{n}
\safemath{\normout}{\gamma}
\safemath{\auxfun}{h}
\safemath{\supp}{\textrm{supp}}%support

\safemath{\indexa}{\ell}
\safemath{\indexb}{r}
\safemath{\indexc}{i}
\safemath{\indexd}{j}

\safemath{\project}{P}%projector

%% file: sec1-intro.tex
% !TEX root = 19SSCL_LAMA.tex

\section{Introduction}\label{sec:intro}

Massive MU-MIMO enables higher per-cell spectral efficiency compared to conventional, small-scale MIMO. 
This improvement, however, comes at a significant increase in baseband processing complexity~\cite{LETM2014}.
In particular, data detection at the base-station (BS) in the massive MU-MIMO uplink is among the most critical tasks in terms of power consumption and throughput~\cite{mimo_overview}.
To exacerbate the situation, the complexity of optimal, maximum a-posteriori (MAP), data detection grows exponentially in the number of user equipment (UE) antennas~\cite{JGMS2015conf}, which prevents its implementation in practice.
% 
% As a result, optimal data detection quickly becomes prohibitive 
% Moreover, data detection in the uplink is among the most critical processing tasks in terms of implementation complexity, power consumption, and throughput in massive MU-MIMO systems \cite{mimo_overview}.
% 

% \fref{fig:sysmodel} illustrates a $B\times U$ massive MU-MIMO uplink system in which~$U$ UEs transmit data simultaneously to a $B$-antenna BS.
%
% Data detection at the BS generates estimates of the UEs' transmit symbols and is among the most critical tasks in terms of throughput, power consumption, and silicon area~\cite{LETM2014}.
% 
% Hence, existing algorithms and corresponding ASIC designs rely either on idealistic channel-hardening assumptions  \cite{TCZ2016,CCTSUY2017} or deploy approximate methods \cite{LETM2014,indiachemp} to reduce complexity.
% 
% As we will show, these solutions result in high error rates (i) under realistic propagation conditions with correlation and per-UE path loss and (ii) if the number of UEs and BS antennas are similar.
% 

To enable high-throughput massive MU-MIMO data detection, a variety of low-complexity algorithms (see, e.g.,~\cite{LETM2014,indiachemp}) and application-specific integrated circuits (ASICs)~\cite{TCZ2016,CCTSUY2017,TPLOZ2018} have been proposed. 
These algorithms and ASIC designs either rely on idealistic channel-hardening assumptions \cite{TCZ2016,CCTSUY2017} or deploy approximations \cite{LETM2014,indiachemp} to reduce complexity.
Unfortunately, both of these simplifications result in high error rates (i) under realistic propagation conditions, such as correlation and per-user path loss, and (ii) in systems where the number of UEs is equal to the number of BS antennas.
As a consequence, achieving near-optimal performance in realistic systems necessitates novel data detection algorithms that can be implemented efficiently.

% scale poorly in massive MU-MIMO uplink when the number of transmit antennas is similar to that of receive antennas at the BS.
% 
% Moreover, these algorithms are designed to operate on channels that have i.i.d. Rayleigh fading, which in general, does not hold in practical massive MU-MIMO systems.
% 
% Therefore, MIMO detection ASICs that provide near-optimal performance that works well in practical channels

% for  (i) under realistic propagation conditions with correlation and per-UE path loss and (ii) if the number /of UEs and BS antennas are similar.
% are the key to successlful deployment of massive MU-MIMO.
% 
% As a result, it is 

\subsubsection*{Contributions}
We propose the first data detector ASIC that achieves near-MAP performance for 32 UEs under realistic propagation conditions. Furthermore, the ASIC provides soft-input soft-output (SISO) capabilities for iterative detection and decoding. 
The algorithm builds upon the large-MIMO approximate message passing (LAMA) algorithm~\cite{JGMS2015conf}, which achieves MAP-optimal error-rate performance for Rayleigh fading channels and in the large-antenna limit,
\revision{
assuming that the UE-to-BS antenna ratio is less than a threshold that depends on the constellation.
% ~\cite{JGMS2015conf}.
% 
In contrast to linear data detectors, LAMA exploits information on the constellation to improve performance; for QPSK, for example, LAMA achieves  optimal performance in the large-antenna limit and for systems where the number of UE and BS antennas are identical.
Since practical systems are finite-dimensional and real-world channels exhibit correlation,  we include algorithm-level optimizations to support realistic channels with LAMA.} 
To achieve high throughput at low area, our ASIC uses coarse-grained pipeline interleaving, processing two detection problems within the same architecture. 
The fabricated 28nm CMOS ASIC outperforms existing designs under realistic channel conditions and for systems in which the number of UEs is comparable to the number of BS antennas.

%% file: sec2-system.tex
% !TEX root = 19SSCL_LAMA.tex

\section{Massive MU-MIMO Data Detection}

%\subsection{System Model}
% 
We consider the uplink of a coded massive MU-MIMO system with $\MT$ single-antenna UEs and~$\MR$ BS antennas. 
The information bit vectors $\bmb$ of the $\MT$ UEs are encoded on a per-UE basis (e.g., using a convolutional code) and the resulting coded bit-stream vectors~$\bmx$ are mapped (using Gray labeling) to a sequence of transmit vectors~$\bms\in\setO^\MT$, where~$\setO$ corresponds to the constellation of size $2^Q$.   
Each transmit vector~$\bms$ is associated with $\MT Q$ binary values $x_{u,q}\in\{0,1\}$, $\mt=1,\ldots,\MT$, $q=1,\ldots,Q$, corresponding to the $q$th bit of the $\mt$th entry (spatial stream) of~$\bms$. 
We assume $\Exop_\bms[\bms\bms^\Herm]=\Es\bI_\MT$, where $\Es$ is the symbol variance. 
The baseband input-output relation of the MU-MIMO channel is modeled as \mbox{$\bmy=\bH\bms+\bmn$}, where~$\bH\in\complexset^{\MR\times\MT}$ is the MIMO channel matrix, $\bmy\in\complexset^{\MR}$ is the received vector at the BS, and~$\bmn$ is $\MR$-dimensional i.i.d.\ zero-mean complex Gaussian distributed noise with variance $\No$ per entry. 
We assume that~$\bH$, $\No$, and $\Es$ are known at the BS.
%\footnote{In practice, channel-state information is acquired through pilot-based training and hence, not perfect. 
% Since imperfect CSI penalizes the performance of all considered MIMO detection algorithms in a similar way,
 % (see, e.g., \cite{Haene08} for details), we  assume---for the sake of simplicity of exposition---perfect CSI throughout the paper.}  

\subsection{Iterative MIMO Decoding}

Iterative detection and decoding in MIMO systems achieves near-optimal spectral efficiency in MIMO wireless systems \cite{studer2011asic}.
% \cs{reference on iterative MIMO decoding; eg hochwald and ten brink, achieving near capacity in a multi-antnena wireless channel}. 
Reliability information on the coded bits, often expressed as log-likelihood ratios (LLRs), is iteratively  exchanged  between the MIMO data detector and the channel decoder. In each iteration, a soft-input soft-output (SISO)-capable MIMO data detector computes extrinsic LLRs for the coded bits $x_{u,q}$ as 
\begin{align*} 
% \label{eq:LLR}
\LD_{u,q} = \log\!\left(\frac{\ProbMacro{x_{u,q}=1\mid\bmy}}{\ProbMacro{x_{u,q}=0\mid\bmy}}\right) -\LA_{u,q},
\end{align*}
using the received vector~$\bmy$ and a-priori LLRs $\Lambda^\text{prior}_{u,q}$, $u=1,\ldots,\MT$, $q=1,\ldots,Q$, obtained from the channel decoder. The extrinsic LLRs $\LD_{u,q}$, which represent reliability estimates for each coded bit $x_{u,q}$, are then passed to the channel decoder, which computes \emph{new} a-priori LLRs $\LA_{u,q}$, $\forall u,q$, that are used by the MIMO data detector in the next iteration. After a small number of iterations $I$, the channel decoder generates final decisions $\hat\bmb$ for the information bit vector~$\bmb$.%\footnote{Early-termination schemes that reduce unnecessary iterations in turbo decoders (e.g., \cite{KP2009}) may also be used in iterative MIMO systems to reduce power consumption or to improve the (average) throughput.}

\setlength{\textfloatsep}{5pt}% Remove \textfloatsep
\begin{algorithm}[tb]
\footnotesize
\caption{Large MIMO AMP (LAMA) Algorithm}\label{alg:LAMA}
\begin{algorithmic}[1]
\STATE {\bf inputs:} $\bH$, $\bmy$, $\No$, and $\Lambda^\text{prior}_{u,q}$, $\forall u,q$
\STATE {\bf preprocessing:} $\tildebG = \bI_{\MT} - \textnormal{diag}(\bG)^{-1}\bG$ with $\bG = \bH^\Herm\bH$,  $\widetilde{\bmy}^\text{MF} = \textnormal{diag}(\bG)^{-1}\bH^\Herm\bmy$, and $g_\mt =G_{\mt\mt} / \MT$, $\mt=1,\ldots,\MT$
\STATE {\bf initialize:} $\bmz^1=\shat^1 =\bm0_{\MT\times1}$, and $\rho^1 = 0$
% compute the soft-symbols $\bms^1$ and variance $\tau^1$ w.r.t $\Lambda^\text{prior}$ and let $\bmalpha = \bm0$, $\bmz^1 =\widetilde{\bmy}^\text{MF} + \tildebG\bms^1 + \bmalpha^1$
% 
\FOR{$t=1,2,\ldots,\tmax$}
\STATE{\bf mean and variance estimation:} 
\STATEx\hspace{1em}
$\shat^{t+1} = \mathsf{F}(\bmz^t, \rho^t\bmg,\bmLprior)$ \hfill (mean update)
\STATEx\hspace{1em}
$\bmtau^{t+1} = \mathsf{G}(\bmz^t, \rho^t\bmg,\bmLprior)$ \hfill (variance update)

\STATEx\hspace{1em}
$\hat{\tau}^{t+1} = \frac{1}{\MR}\bmg^\Tran \bmtau^{t+1}$
\STATEx\hspace{1em}
$\bmalpha^{t} = \bmz^{t} - \shat^t$ \hfill(Onsager term)
\STATEx\hspace{1em}
$b^{t} = \rho^t \hat{\tau}^{t+1}$

\STATE {\bf interference cancellation:}

\STATEx\hspace{1em}
$\bmz^{t+1} = \tildebmymf + \tildebG \shat^{t+1} + b^t \bmalpha^t$\hfill (interference cancellation)
\STATEx\hspace{1em}
$\rho^{t+1} = (\frac{1}{B}\No + \hat{\tau}^{t+1})^{-1}$ \hfill (post-equalization SINR update)

\ENDFOR

\STATE {\bf output:} extrinsic LLR values $\Lambda^\text{d}_{u,q}$, $\forall u=1,\ldots,\MT,q=1,\ldots,Q$
\end{algorithmic}
\end{algorithm}

\subsection{Hardware Friendly LAMA Algorithm }
% 
% In this section, we quickly summarize the LAMA algorithm.
% 
LAMA is an efficient data detection algorithm based on approximate message passing (AMP), that is provably optimal (in terms of error-rate performance) in the large-system limit (i.e., fix $\beta=B/U$ and  $B\to\infty$) with i.i.d.\ Rayleigh fading channels~\cite{JGMS2015conf}.
In each of its $\tmax$ iterations, LAMA decouples the MIMO system into parallel and independent AWGN channels with equal signal-to-interference-plus-noise ratio (SINR). 
As a result, LAMA optimally denoises the parallel AWGN channels in every iteration, which successively increases the post-equalization SINR and improves the error-rate performance.
% Once done, LAMA returns the decoupled signal estimate for detection, which can be done either via hard decisions or soft estimates via log likelihood ratios. 

To deal with realistic channel conditions (such as correlation and per-UE path loss), we apply algorithm-level modifications to the original LAMA algorithm in~\cite{JGMS2015conf}. 
First, we transform LAMA so that it operates on the $\MT\times\MT$ dimensional Gram matrix $\bG=\bH^\Herm\bH$, instead of the $\MR\times\MT$ channel matrix $\bH$, which reduces the per-iteration complexity. 
Second, we deploy message damping techniques~\cite{RSF2014} to reduce the performance loss of LAMA in finite-dimensional systems that exhibit correlation and large-scale UE fading. Specifically, we damp the updates of $\hat\tau^t$ and~$\rho^t$ by a  factor $\theta\in(0,1]$\revision{, i.e., we use $\hat\tau^t_\text{d}$ instead of $\hat\tau^t$ in line 6 of \fref{alg:LAMA}, and $\hat\tau^t_\text{d} = \theta \hat\tau^t + (1-\theta) \hat\tau^{t-1}_\text{d}$ }. 
%\oc{Where is this $\theta$ in Algorithm 1?}
%
Third, we include support for iterative detection and decoding. 
\revision{The implemented LAMA algorithm is summarized in \fref{alg:LAMA} (message damping details are excluded).}
The functions $\mathsf{F}$ and $\mathsf{G}$ correspond to the posterior mean and variance applied element-wise,
 \revision{i.e., $\mathsf{F}(z,\rho,\Lambda^\text{prior}) = \Exop_S (S\vert z = S + \rho^{-1/2}N)$, $N\sim\setC\setN(0,1)$ and $p(S)$ can be derived from the a-priori LLRs~$\Lambda^\text{prior}$; $\mathsf{G}$ can be derived similarly}---see~\cite{JGMS2015conf} for the details.

%% file: sec4-vlsi.tex
\setlength{\textfloatsep}{5pt}% Remove \textfloatsep
\begin{figure}
\centering
\includegraphics[width=0.8\columnwidth]{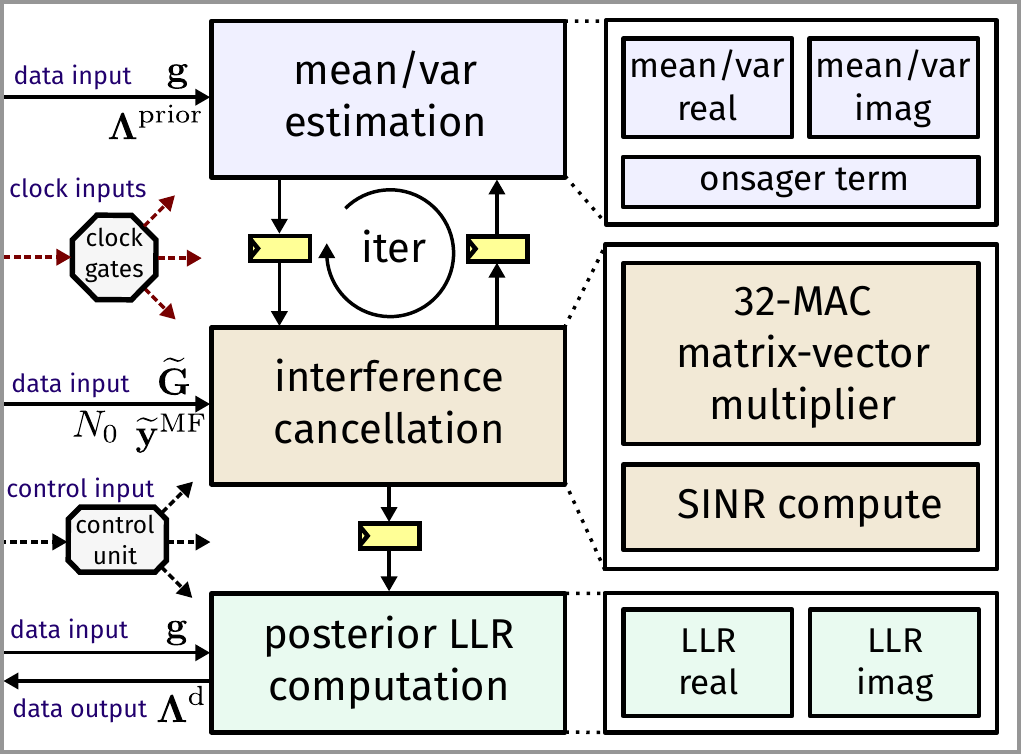}
\vspace{-0.15cm}
\caption{Top-level architecture of the LAMA data detector. Data detection is carried out in a pipeline-interleaved manner, which iteratively processes two independent detection problems in the mean and variance (MV) estimation unit and the interference cancellation (IC) unit.}
\label{fig:blk_detailed}
\end{figure}

\section{VLSI architecture}

\setlength{\textfloatsep}{5pt}% Remove \textfloatsep
\begin{figure*}
\noindent\begin{minipage}[t]{0.39\textwidth}
\centering
\begin{figure}[H]
\centering
% \subfigure[]{
\includegraphics[height=5.0cm]{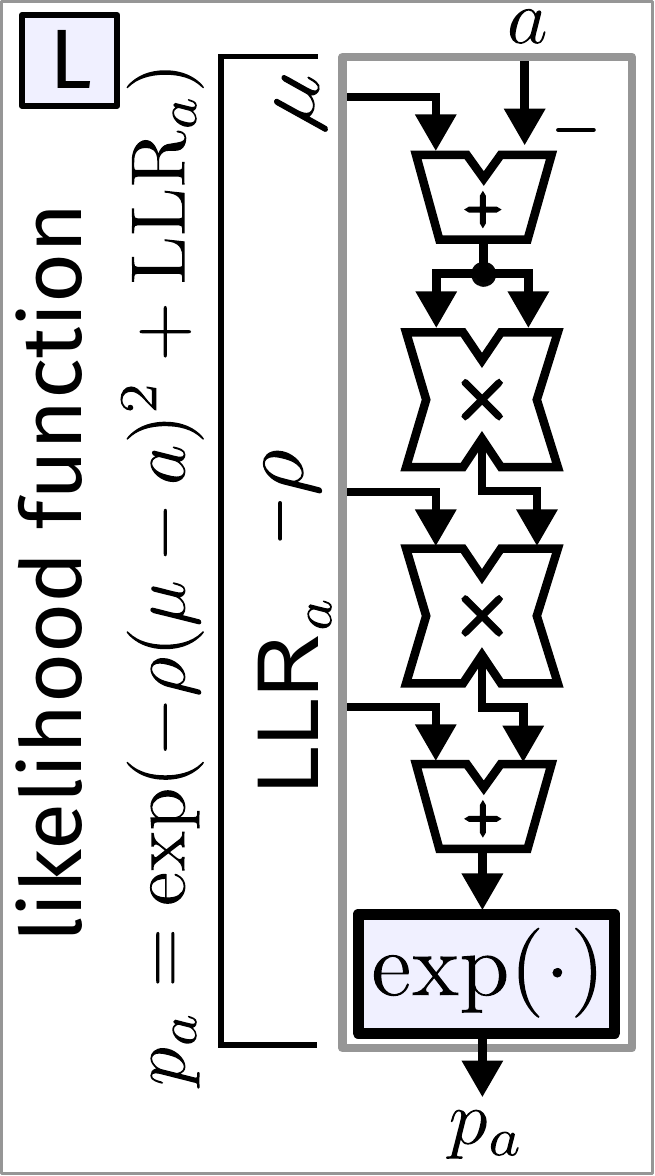}
% }
% \hspace{1cm}
\hfill
% \subfigure[Cannon's algorithm \cite{cannon}]{
\includegraphics[height=5.0cm]{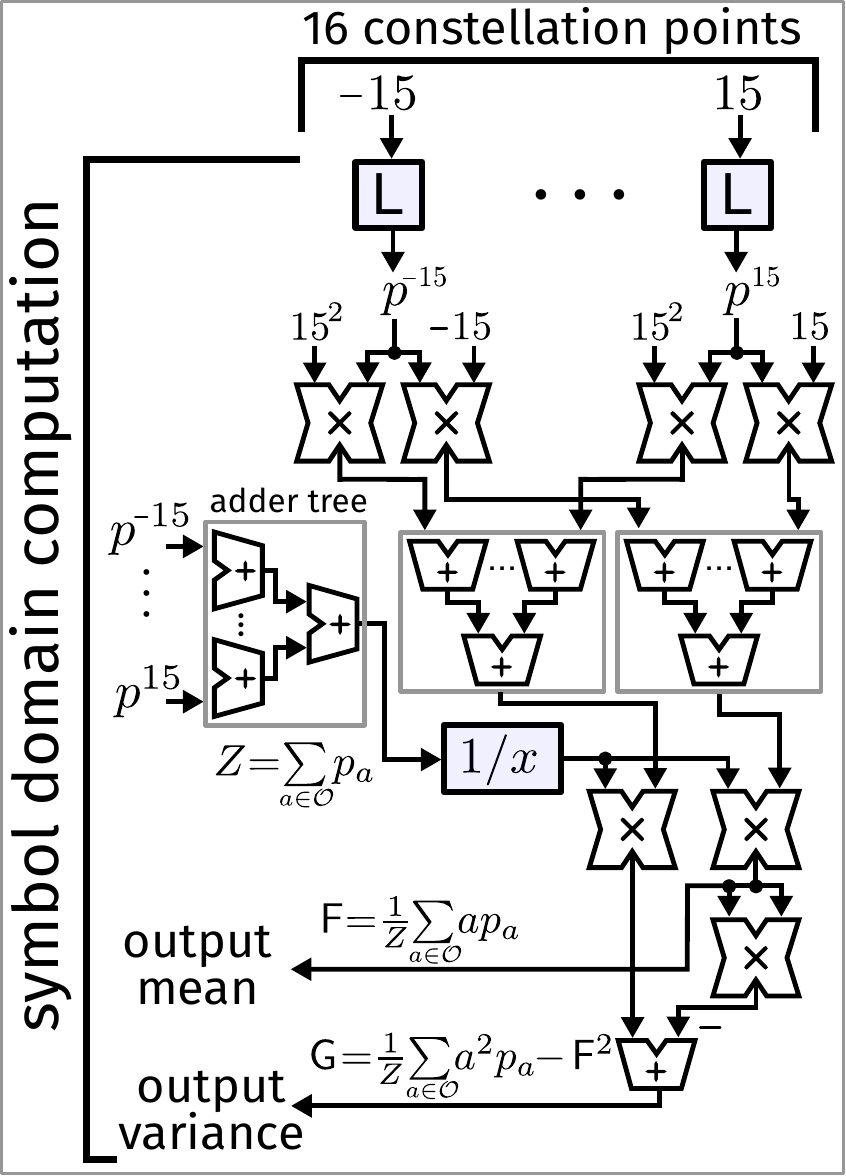}
\caption{Original MV unit for 256-QAM with separation into two 16-PAM units.
 % mean and variance in parallel.
 Exact mean and variance computation entails high complexity and requires high arithmetic precision.
% The input is circularly shifted in the shift registers
}
\label{fig:F_original}
\end{figure}
\end{minipage}
\hfill
\noindent\begin{minipage}[t]{0.39\textwidth}
\centering
\begin{figure}[H]
\centering
% \subfigure[]{
\includegraphics[height=5.0cm]{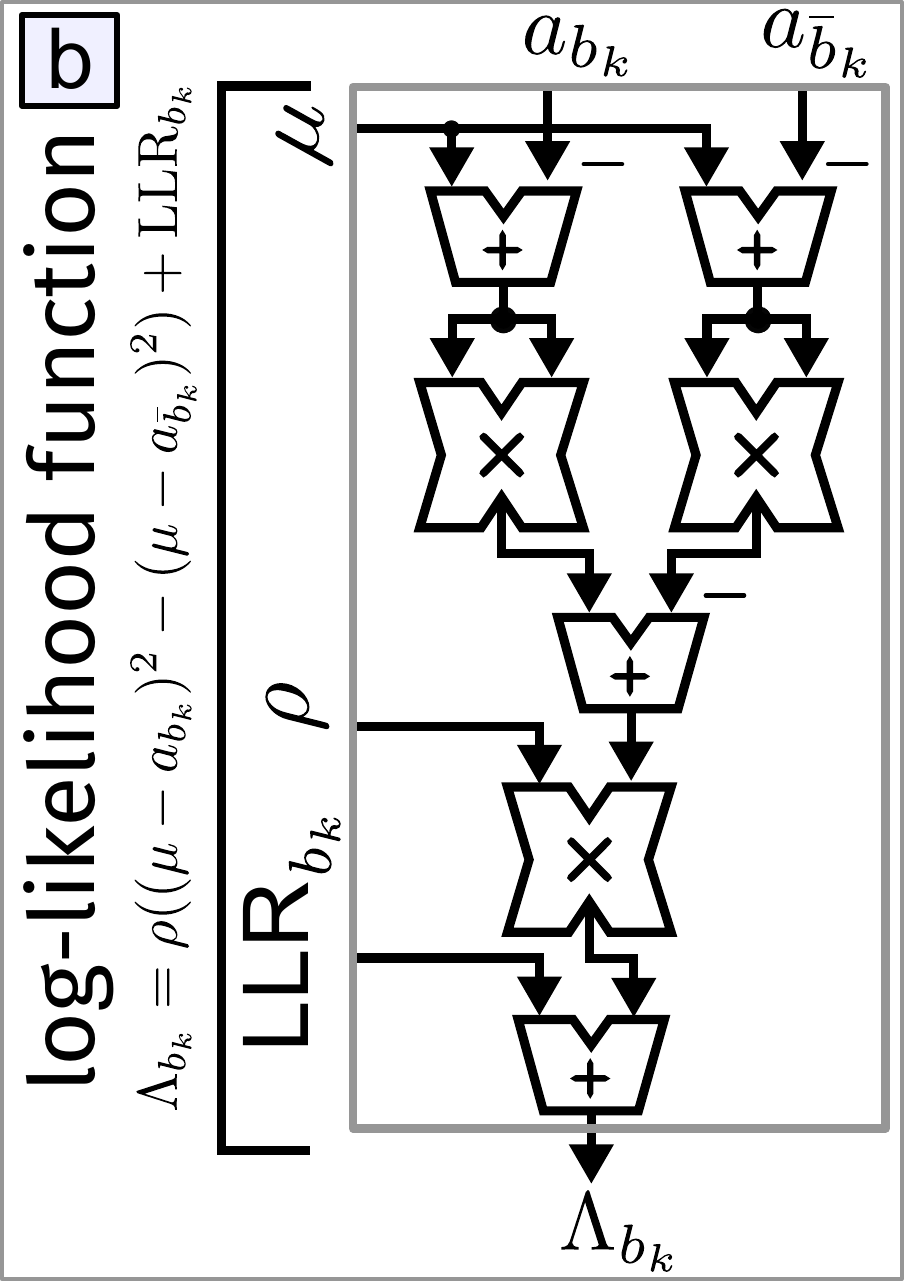}
\hfill
\includegraphics[height=5.0cm]{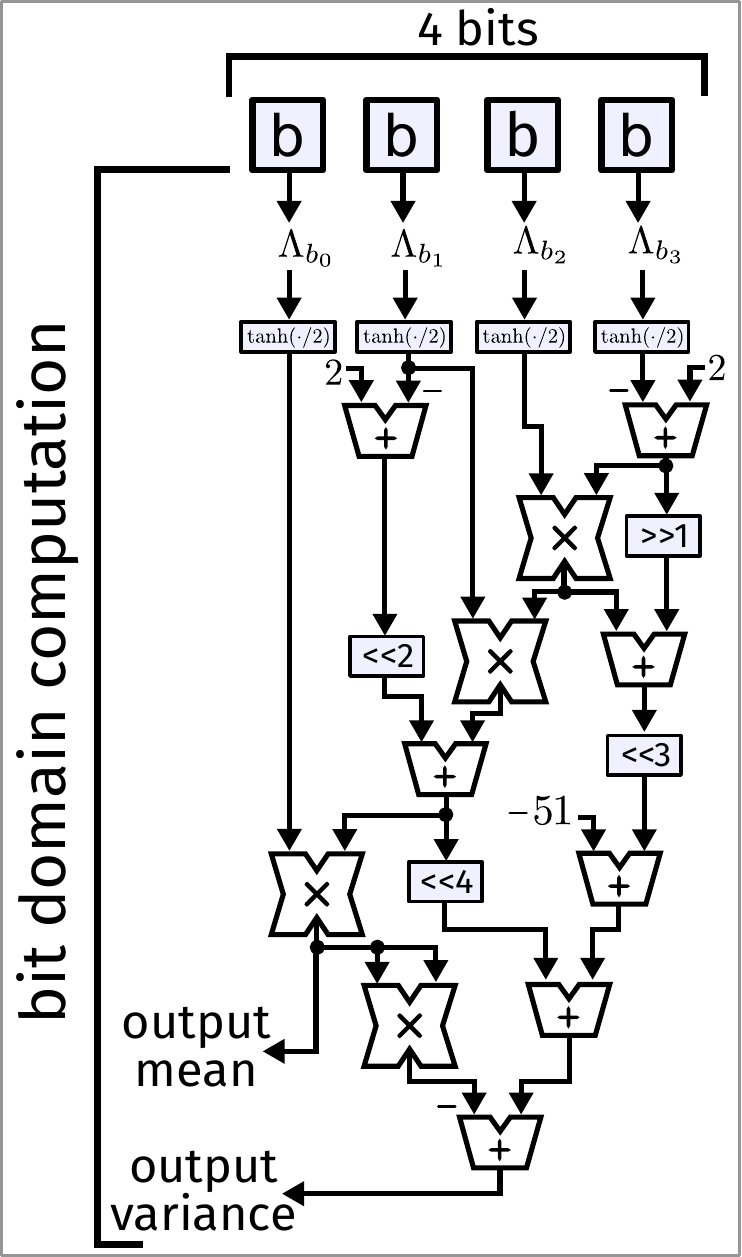}
% }
% \hspace{1cm}
% \subfigure[Cannon's algorithm \cite{cannon}]{
% \includegraphics[height=5.2cm]{figs/F2_likelihood.pdf}
% \includegraphics[height=4.8cm]{figs/cannon.pdf}
\caption{Proposed low-complexity computation of message mean and variance. 
We transform all computations into the bit-domain and use the max-log approximation.
 % and compute the mean and variance estimates in bit-domain.
}
\label{fig:F_new}
\end{figure}
\end{minipage}
\hfill
\noindent\begin{minipage}[t]{0.19\textwidth}
\centering
\begin{figure}[H]
\centering
% \subfigure[straightforward implementation]{\includegraphics[height=5cm]{figs/Fsym.pdf}}
% \hspace{-0.4cm}
\includegraphics[height=5.0cm]{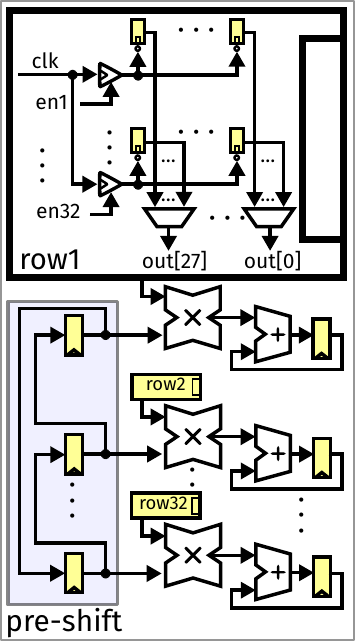}
\caption{
Fan-out reduction of matrix-vector multiplication using Cannon's algorithm~\cite{CannonThesis}.
% We reduce the fan-out of the 32-MAC array's drivers with Cannon's algorithm \cite{cannon}.
% Cannon's algorithm \cite{cannon} is used to reduce the fan-in for 32-MAC array
% , minimizing the ASIC's critical path. 
% The input is circularly shifted in the shift registers
}
\label{fig:cannon}
\end{figure}
\end{minipage}
% \twocolumn
\vspace{-0.5cm}
\end{figure*}

%\subsection{Architecture Overview}

\fref{fig:blk_detailed} depicts the top-level architecture of the LAMA data detector. 
LAMA performs two main tasks per iteration: The first task 
% \cs{these two things should be explained in the algorithm description above} 
estimates the mean and variance (MV) of the data transmitted by each UE; the second task cancels interference (IC) among the UEs---both of these tasks are detailed below. To maximize throughput, two independent detection problems are processed simultaneously in a pipeline-interleaved manner, i.e., one problem per task. 
The two main processing units, namely MV and IC, perform the assigned computations in $\Ts$ clock cycles, and the results of both units are exchanged for further processing in the subsequent iteration. 
In the last $\tmax$ iteration, the outputs from the IC unit are sent to the LLR computation unit, which takes $\TLLR$ clock cycles. Thus, LAMA delivers a new set of $\MT Q$ LLR values at a sustained throughput of
\begin{align}
\label{eq:throughput}
\Theta =  
% \textstyle
\frac{\MT Q}{\tmax\Ts + \TLLR}f_\textnormal{clk} \quad \text{[bit/s]}.
\end{align}
The final design supports $32$ UEs, which requires $\Ts = \TsNumber$ clock cycles and $\TLLR=1$ clock cycle.

\subsection{Mean and Variance Estimation (MV) Unit}

In the first task (line 5 in \fref{alg:LAMA}), the MV unit receives estimates of the UE's data and the associated SINR to compute mean and variance values. 
As shown in \fref{fig:F_original}, a straightforward MV unit would require a large number of multipliers.
Although statistical independence in the real and imaginary parts of the transmitted constellation points simplifies computation from $M^2$-QAM to two $M$-PAM constellations, mean and variance computation for $16$-PAM (to support 256-QAM) still requires $16$ likelihood function units and a division, resulting in high complexity.
Furthermore, the intrinsic LLR values obtained from the channel decoder must be transformed from bit-domain to symbol-domain for SISO processing.
To reduce complexity, existing ASICs~\cite{TCZ2016,TPLOZ2018,CCTSUY2017} use hard-symbol clipping or linear approximations, and do not provide support for SISO processing.
However, accurate message mean and variance computation is key to support realistic channels and systems with a comparable number of UEs and BS antennas.

To accurately compute the message mean and variance at low complexity, we (i) compute all quantities in the bit-domain, (ii) exploit Gray-mapping symmetries, and (iii) use the max-log approximation. 
\revision{
The conversion into the bit-domain and the max-log approximation only requires $4$ log-likelihood functions for $16$-PAM, instead of $16$ functions in the symbol domain.
In addition, we avoid the need of a division per UE by using a LUT-based $\text{tanh}(\cdot)$ function as in~\cite{studer2011asic} with 7 input bits.
}

The resulting architecture, depicted in \fref{fig:F_new}, also avoids the need of a division per UE. 
% 
% 
% \oc{On the left side of \fref{fig:F_new}, the formula shows $\rho$ multiplying, but the circuit has $-\rho$. Also, in the entries of the first adders, the connection from $\mu$ to the second adder is missing.}
% 
%
Furthermore, the architecture naturally supports SISO processing for iterative MIMO decoding.
Our simulations in \revision{\fref{sec:PERperf}} for various antenna configurations  and channel models show that our approach entails a negligible performance loss at around $4\times$ lower area.
% 

% \fref{fig:F_new} depicts the proposed method which reduces from $16$ to $4$ real-valued multiplications per UE, and the required precision without noticeable error-rate performance loss.
% 
% 
% The MV unit also includes logic to process soft-inputs in the form of LLR values.
% computations without any increased overhead.
% 

\subsection{Interference Cancellation (IC) Unit}
% 
% \cs{this can definitely be streamlined.}
%
In the second task (line 6 in \fref{alg:LAMA}), the IC unit performs interference cancellation and  updates the SINR. 
\subsubsection{$32$-MAC matrix-vector multiplication}
Interference cancellation requires a $32\times32$ complex-valued  matrix-vector multiplication, which we compute sequentially in $32$ clock cycles using a linear array of $32$ complex-valued multiply-accumulate (MAC) units in a column-by-column fashion.
To minimize the critical path caused by the large fan-out of a conventional linear array of MAC units, \revision{\fref{fig:cannon}} shows a simplified version of Cannon's algorithm~\cite{CannonThesis}, which circularly shifts the array's input vector while sequentially processing rows of the matrix \revision{over multiple clock cycles}; this reduces the vector memory fan-out from 32 MAC units to one MAC unit and a register. 
To further reduce the critical path and simplify placement, each row of the Gram matrix is stored next to each MAC unit with standard-cell-based latch-arrays.
\subsubsection{SINR computation}

% \cs{I would shorten the discussion of the NR divider but rather add a few sentences about fixed point performance; for the NR architecture you can refer to the mmse pic paper anyway}

The post-equalization SINR is computed in parallel using a Newton-Raphson (NR) reciprocal unit~\cite{studer2011asic}. 
We first shift the input $x$ according to $\bar x=2^\alpha x$, $\alpha\in\mathbb{Z}$ so that $\bar x\in[0.5,1)$, resulting in  high numerical stability.
% 
% This normalization results in $\bar y = \bar x^{-1}\in (1,2]$, and thus, the subsequent NR iterations are carried out with high numerical stability. 
% 
Based on an initial guess obtained from a look-up table, a single NR iteration is sufficient to compute $\bar{y}_1\simeq \bar{x}^{-1}$; the final result $y = 2^\alpha\bar{y}_1 $ corresponds to an approximation of $x^{-1}$.
% 
% Based on an initial guess of $\bar y_0$ obtained from a look-up table, a single NR iteration $\bar{y}_1 = 2 \bar{y}_0 - \bar{x} \bar{y}_0^2$ is sufficient; the final result $\bar{y}_1$ and hence, $y = 2^\alpha\bar{y}_1 $, corresponds to an approximation of $\bar{x}^{-1}$ and $x^{-1}$, respectively.
% 

%% file: sec5-implementation.tex
% !TEX root = 19SSCL_LAMA.tex

%%%%%%%%%%%%%%%%%%%%%%%%%%%%%%%%%% 
% new section
% %%%%%%%%%%%%%%%%%%%%%%%%%%%%%%%%%%%%
\section{Implementation Results and Comparison}\label{sec:Implementation}

% \subsection{\todo{ASIC Fixed-point parameters}}

\begin{figure}[t]
\centering
\subfigure[i.i.d. Rayleigh]{\includegraphics[height=4.45cm]{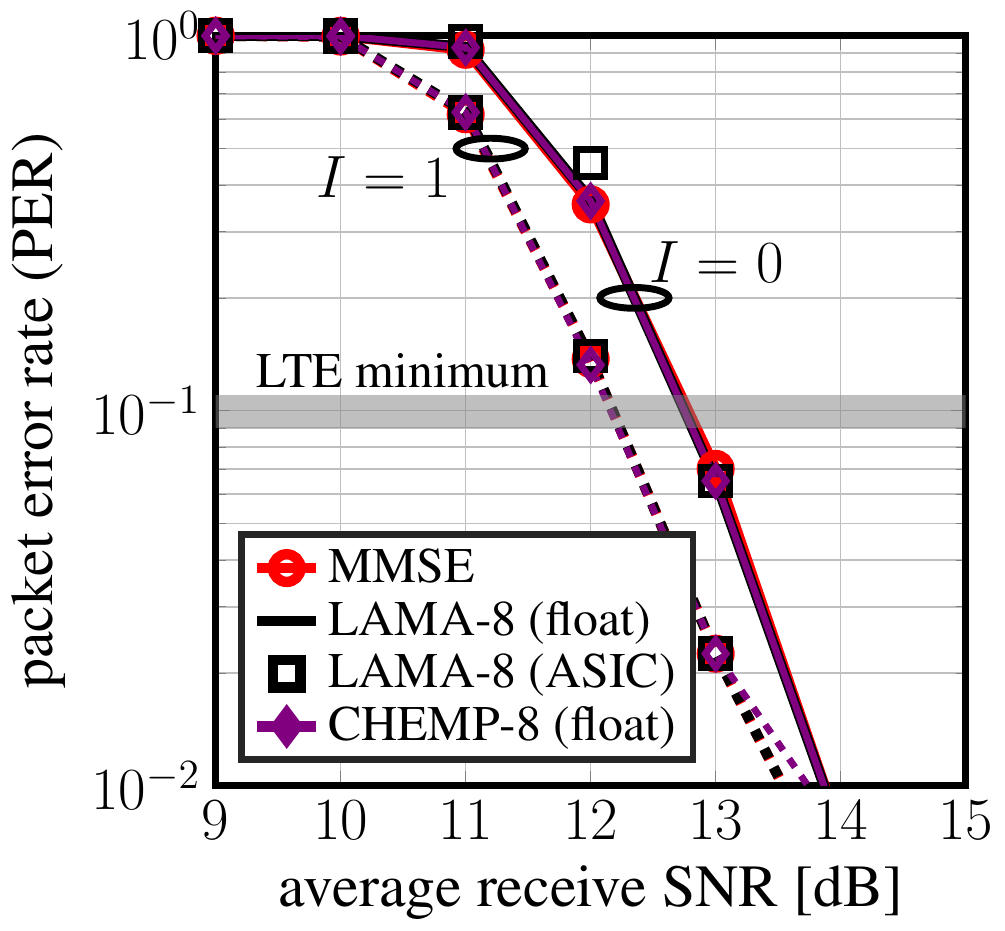}}
\hfill
% \hspace{-0.2cm}
\subfigure[Typical Urban Micro]{\includegraphics[height=4.35cm]{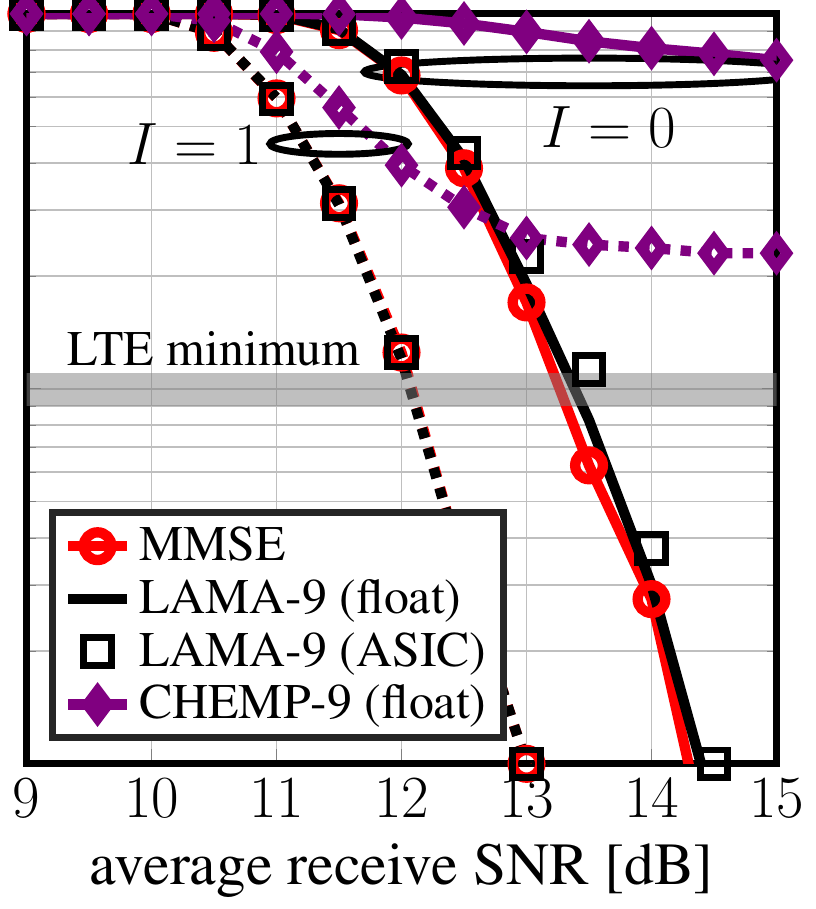}}
\vspace{-0.4cm}
\caption{$256\times32$ massive MU-MIMO; $R=0.5$; 256-QAM\revision{; 9600 bits$/$packet.}}
\label{fig:256x32}
\vspace{0.1cm}
\centering
\subfigure[i.i.d. Rayleigh]{\includegraphics[height=4.45cm]{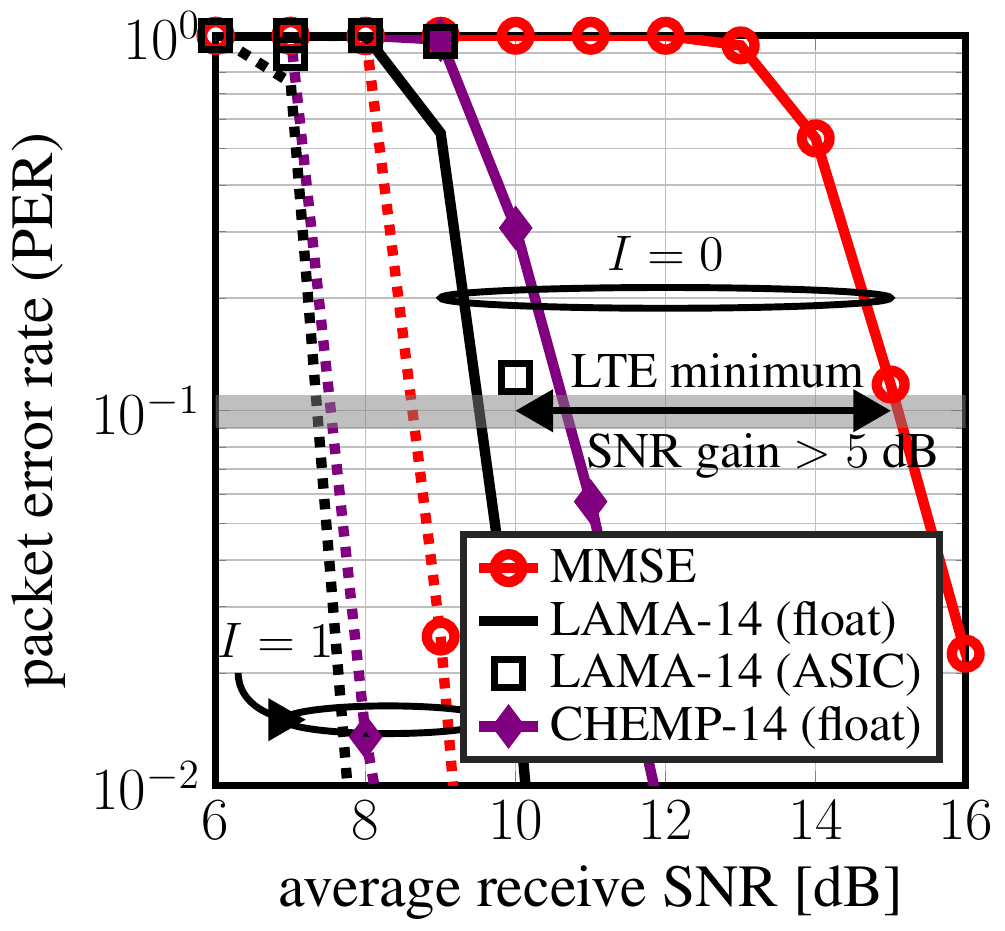}}
\hfill
\subfigure[Typical Urban Micro]{\includegraphics[height=4.35cm]{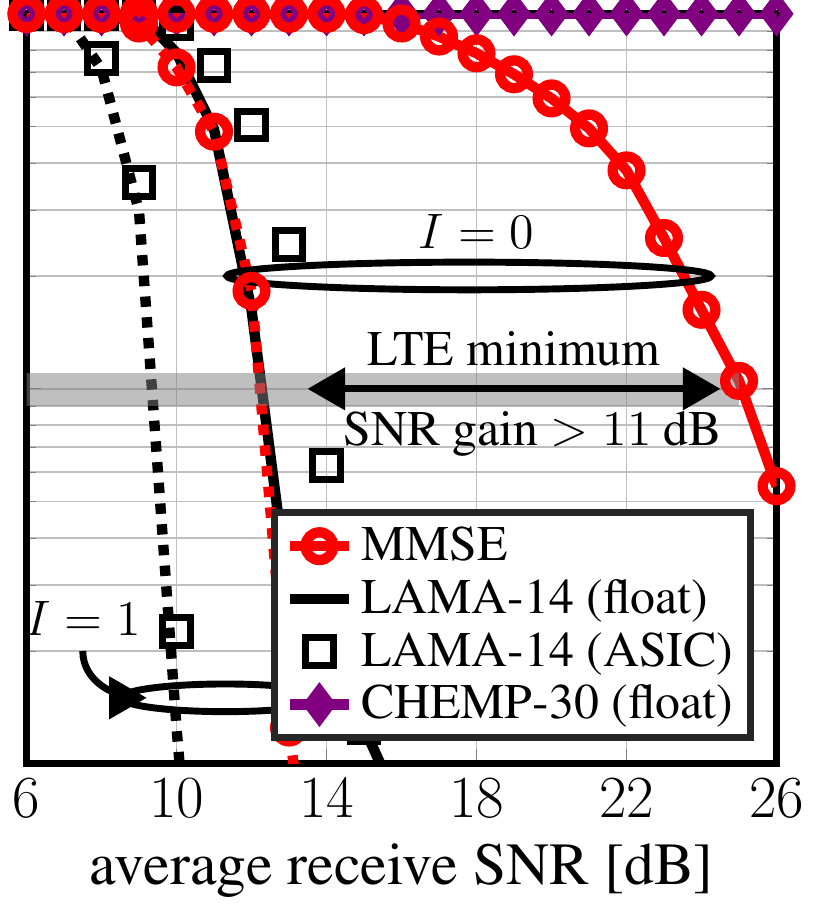}}
\vspace{-0.4cm}
\caption{$32\times32$ massive MU-MIMO; $R=0.75$; QPSK\revision{; 3600 bits$/$packet.}}
\label{fig:32x32}
\end{figure}

%\subsection{Packet Error-Rate (PER) Performance}
\label{sec:PERperf}
Figs.~\ref{fig:256x32} and \ref{fig:32x32} show the PER of our LAMA ASIC in comparison with the linear minimum-mean squared-error (MMSE) equalizer and channel hardening-exploiting message passing (CHEMP) algorithm~\cite{indiachemp}. 
\revision{The number of algorithm iterations are indicated after the dash; e.g., LAMA$-14$ represents LAMA with 14 iterations. Outer iterations over the channel decoder are shown as either none ($I=0$; solid lines) or one ($I=1$;  dashed lines) iteration.
}
\revision{We simulate an LTE-based massive MU-MIMO-OFDM system at $f_c=2$\,GHz with 1200 active subcarriers and per-user convolutional coding with rate~$R$.}
\revision{We use two channel models: (a) Rayleigh fading and (b) WINNER II typical urban micro \cite{WINNER} to model a realistic propagation environment.}
For a typical $256\times32$ ($\MR\times\MT$) massive MU-MIMO scenario, LAMA achieves the same performance as linear MMSE, but avoids a matrix inversion; CHEMP suffers an error floor above 10$\%$ PER.
For the challenging $32\times32$ system, LAMA significantly outperforms the linear MMSE detector, achieving more than $11$\,dB SNR improvements for the typical urban micro channel; CHEMP fails to successfully detect packets. 
\revision{Extensive numerical simulations have been carried out to determine the ASIC's fixed-point parameters; the implemented design achieves near-floating-point performance.}
% 
% We also note that the LAMA fixed-point ASIC suffers a negligible performance loss compared to a MATLAB floating-point implementation. 

% A prototype ASIC has been fabricated in 28nm CMOS; 
\subsection{Implementation Results}
\begin{figure}
\centering
\noindent\begin{minipage}{0.52\columnwidth}
% \subfigure[Chip micrograph; LAMA data detector ASIC is highlighted.]{

\includegraphics[height=4.0cm]{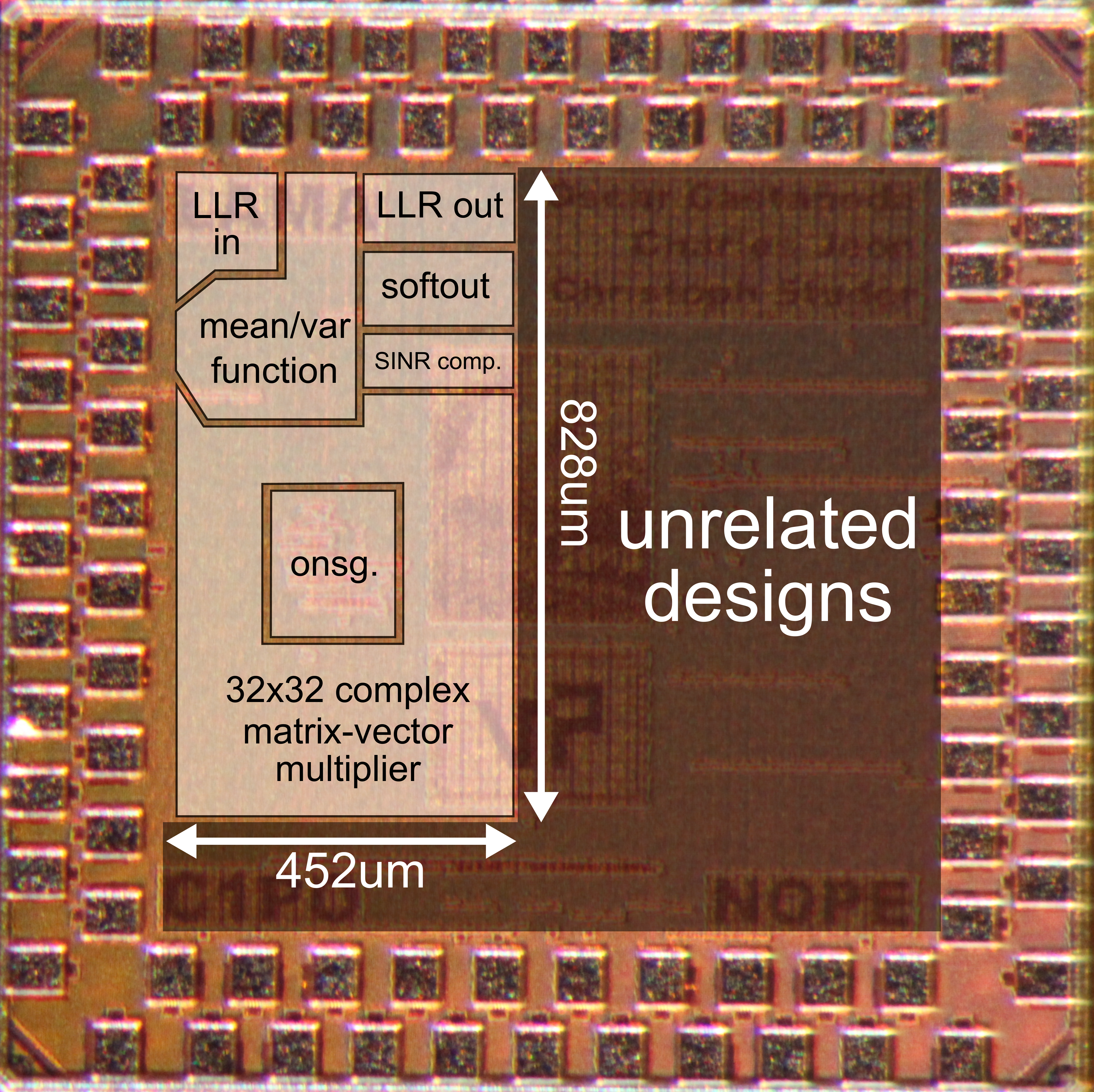}
\vspace{-0.1cm}
\caption[caption]{Chip micrograph; LAMA\\\hspace{\textwidth}data detector ASIC is highlighted.}
\label{fig:chip}
% }
\vspace{-0.3cm}
\end{minipage}
\noindent\begin{minipage}{0.45\columnwidth}
% \subfigure[Measured frequency and energy for different core voltages.]{
\centering
\includegraphics[height=4.0cm]{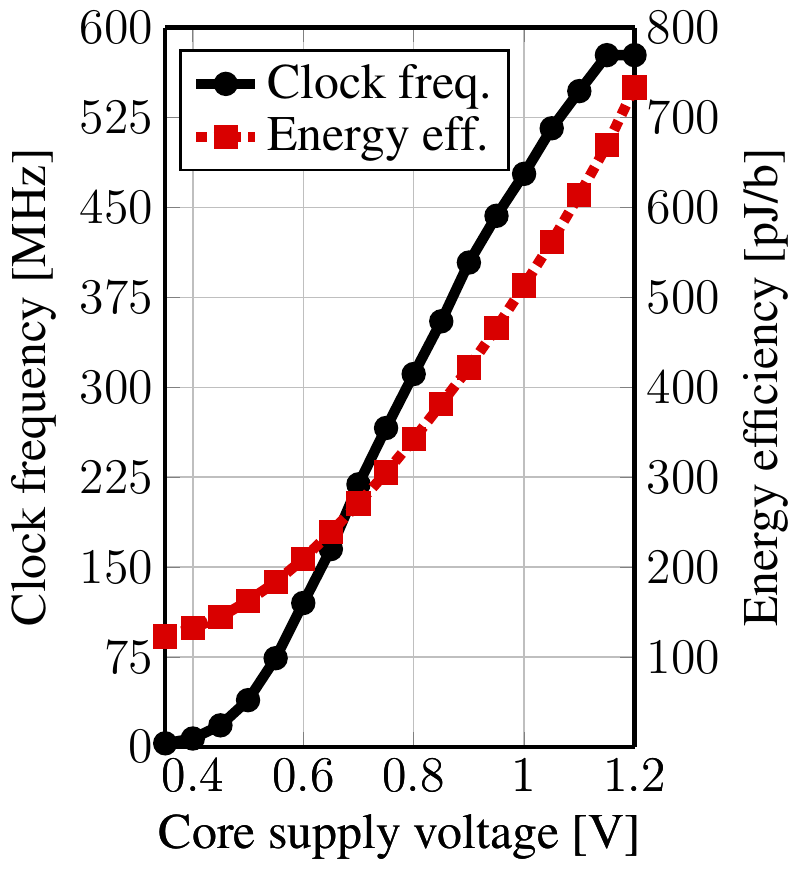}
\vspace{-0.1cm}
\caption{Measured frequency and energy for different core voltages.}
\label{fig:measurements}
% }
\vspace{-0.3cm}
\end{minipage}
\end{figure}

\fref{fig:chip} shows a micrograph of the fabricated and fully-functional 28nm CMOS ASIC with the LAMA detector core highlighted. The LAMA ASIC only occupies $0.37$\,mm$^2$; the rest of the chip contains  unrelated designs.
\revision{The clock signal was generated by a VLSI test system and directly fed into the ASIC.}
%
%The total chip area is $2$\,mm$^2$ \oc{I would not mention the 2mm2 area; and also I would change the image to read ``unrelated designs''; the other things have nothing to do and we should be clear about that!} but the LAMA detector only occupies $0.37$\,mm$^2$.
%
At nominal supply of $0.9$\,V at $300$\,K, the ASIC reaches a maximum measured clock frequency of $400$\,MHz at $151$\,mW, which  results in $354$\,Mb/s for $32$ UEs transmitting $256$-QAM. 
% \oc{This throughput is for 8 iters, which is what is used in Rayleigh. Should we stick with this number or use the one for urban micro (315Mbps , with 9 iters)?}. 
 
\fref{fig:measurements} shows measured energy-efficiency in pJ/bit obtained via voltage-frequency scaling. By reducing the supply close to the threshold voltage, the detector achieves optimal energy-efficiency: at $0.35$\,V we have $123$\,pJ/bit (achieving $2.66$\,Mb/s). If maximum throughput is desired, one can increase the supply to $1.15$\,V and obtain $511$\,Mb/s (at $670$\,pJ/b efficiency). 

\fref{tbl:comparison} compares LAMA to state-of-the-art massive MU-MIMO data detectors. 
Our LAMA ASIC achieves more than $4\times$ improved normalized area efficiency than \cite{TPLOZ2018}, which computes a matrix inversion. 
Although LAMA achieves lower area efficiency (in Gb/s/$\text{mm}^2$) than the detectors in \cite{TCZ2016,CCTSUY2017}, these  designs suffer an error floor higher than LTE specifications under realistic channel conditions (cf.~Figs~\ref{fig:256x32} and~\ref{fig:32x32}). 
We note that the nominal energy efficiency is inferior to other designs due to increased arithmetic precision requirements in support of realistic channel conditions and symmetric massive MU-MIMO systems.
To the best of our knowledge, the proposed LAMA ASIC is the first silicon prototype of a $32$-UE massive MU-MIMO data detector that provides near-optimal error rates under realistic  propagation conditions and for symmetric systems. 
Both of these advantages are critical to BS providers as one can support up to 32 UEs with relatively small ($B \geq 32$) BS antenna arrays under realistic channel conditions.

% \vspace{-0.5cm}

% 

% \noindent\begin{minipage}{0.4\textwidth}
% \begin{center}  
% \vspace{-0.5cm}
% \captionof{table}{Performance summary and ASIC comparison }
% \label{tbl:comparison} 
% \footnotesize
\begin{table}
\caption{
Performance summary and ASIC comparison}
\footnotesize
\vspace{-0.3cm}
\begin{center}
\scalebox{0.95}{
\renewcommand{\arraystretch}{1.0}    
\begin{tabular}{@{}p{2.3cm} c | c P{1.15cm} P{1.2cm} @{}} 
\toprule[0.2em]
\multirow{2}{4em}{}
& 
% \multicolumn{2}{c|}{This work}
This work  & Tang \cite{TCZ2016} & Chen \cite{CCTSUY2017}   & Tang \cite{TPLOZ2018} 
% & Jeon \cite{JCGS2019} 
\\
\midrule[0.15em]
Max UEs  & $32$ & $32$  & $8$ & $16$ 
% & $16$
\\ 
Algorithm & LAMA  & CHEMP & CHEMP & EPD$^a$ 
% & NOPE$^b$
\\
Soft-in soft-out & {\bf yes}  & no & no$^b$  & no 
% & no 
\\
Modulation & 256-QAM & 256-QAM & QPSK & 256-QAM \\
% & 256-QAM \\
Realistic channels
 & {\bf yes}  & no & no & {\bf yes} \\
 % & {\bf yes}\\ 
\midrule[0.15em]
% \hline
Technology [nm] & 28  & 40& 40 & 28 \\
% & 28\\
% 
Supply [V] & 0.9 & 0.9 & 0.9 & 1.0 \\
% & 0.9 \\
% 
Area [$\text{mm}^2$] & 0.37 & 0.58 & 0.076 & 2.0 \\
% & 0.43\\
% \midrule
Frequency [MHz] & 400 & 425& 500 & 569 \\
% & 380\\
% & 300& 425 & 680 & 560 & 500  & 569 \\ 
% 
Power [mW] & 151 & 220.6& 77.9 & 127 \\
% & 158\\
% 
Throughput [Gb/s]\!\!\!\!  & 0.354 & 2.76& 8 & 1.80\\
 % & 0.219\\
% 
\midrule[0.1em]
Energy$^\textit{c}$ [pJ/b]& 426 &  79.9 & 9.74 & 70.56\\
 % & 720 \\
% \hline
Area~Eff.$^\textit{d}$~[Gb/s/$\text{mm}^2$]\!\!\!\!\! &  0.95 & 4.76 & 105.26& 0.90\\
 % & 0.50\\
% \hline
\midrule[0.1em]
Norm.~Energy$^\textit{c,e,f}$~[pJ/b]\!\!\!\! & 426 & 39.16 & 76.33  & 215 \\
% &  2878  \\
% \hline
Norm.~Area~Eff.$^\textit{d,e,f}$ [Gb/s/$\text{mm}^2$]& {\bf \multirow{2}{*}{0.95}} & \multirow{2}{*}{13.87} & \multirow{2}{*}{19.18}  & \multirow{2}{*}{0.21}  \\
% & {\multirow{2}{*}{0.13}} \\
\bottomrule[0.2em]
\end{tabular}} \\[0.2cm]
\end{center}
% \footnotesize
% 
{\footnotesize $^a$expectation-propagation,
% $^b$non-parametric equalizer,
$^\textit{b}$soft-output support only, 
$^\textit{c}$energy efficiency is $\text{power}/\text{throughput}$,
$^\textit{d}$area efficiency is $\text{throughput}/\text{area}$,
$^\textit{e}$technology normalized to 28nm, $V_\text{dd}=0.9\,\text{V}$,
$^\textit{f}$normalized by $(U/32)^2$.}
 % $f_\text{clk}\sim s$, $A\simeq 1/s^2$, and $P_\text{dyn}\sim (1/s) (V_\text{dd}/V_\text{dd}')$
 % $^b$standard technology scaling rules apply; $^c$the ZF mode does not require any parameters.
\label{tbl:comparison} 
\end{table}
% % 
% \end{center}
% \end{minipage}

%% file: secc-appendix.tex
% !TEX root = 19SSCL_LAMA.tex